# Pushing the Limit of high-Q Mode of a Single Subwavelength Dielectric Nanocavity

Lujun Huang[1§], Lei Xu[1§], Mohsen Rahmani[2], Dragomir Neshev[2], and Andrey E Miroshnichenko[1*]

[1]School of Engineering and Information Technology, University of New South Wales, ACT, Canberra, Australia, 2602;

[2]Research School of Physics and Engineering, Nonlinear Physics Centre, The Australia National University, ACT, Canberra, Australia, 2602

[§] These authors contributed equally.

[*] To whom correspondence should be addressed

E-mail: andrey.miroshnichenko@unsw.edu.au

Abstract

High index dielectric nanostructure supports different types of resonant modes. However, it is very challenging to achieve high-Q factor in a single *subwavelength* dielectric nanoresonator due to non-hermtian property of the open system. Here, we present a universal approach of finding out a series of high-Q resonant modes in a single nonspherical dielectric nanocavity by exploring *quasi-bound state in the continuum*. Unlike conventional method relying on heavy computation (ie, frequency scanning by FDTD), our approach is built upon *leaky mode engineering*, through which many high-Q modes can be easily achieved by constructing avoid-crossing (or crossing) of the eigenvalue for pair leaky modes. The Q-factor can be up to $2.3\times10^4$ for square subwavelength nanowire (NW) (n=4), which is *64 times* larger than the highest Q-factor (Q≈360) reported so far in single subwavelength nanodisk. Such high-Q modes can be attributed to suppressed radiation in the corresponding eigenchannels and simultaneously quenched electric(magnetic) at momentum space. As a proof of concept, we experimentally demonstrate the emergence of the high-Q resonant modes (Q≈380) in the scattering spectrum of a single silicon subwavelength nanowire.

The Q-factor of a cavity is defined as the energy dissipation per unit circle versus energy stored in the resonator. Generally, it is desirable to have a high-Q factor for an optical resonance since it can achieve extreme energy confinement that can significantly reduce the threshold of lasing and enhance the light-matter interaction. The most widely used methods of realizing high-Q modes are built upon the photonic crystal cavity or whispery gallery cavity[1,2]. However, the on-chip lasing source requires high-Q resonator being of the subwavelength scale. Recently, subwavelength high-index dielectric nanostructure has emerged as a promising platform to realize CMOS-compatible nanophotonics since it supports Mie-type resonances (also known as leaky mode resonance) with reduced dissipation loss[3,4]. The value of the Q-factor for leaky mode resonances, however, is limited in a subwavelength dielectric resonator due to the non-Hermicity of open system. Up to date, there is a fundamental question remained unanswered: *how to design a dielectric particle with the largest Q-factor?* For practical applications, it would be essential to have a method of realizing extremely high-Q modes in arbitrary subwavelength dielectric nanocavities.

In this letter, we report the general rule of thumb to find such high-Q mode in a single nonspherical nanocavity (i.e. rectangular NW, cuboid, and disk). We demonstrate our analysis based on a rectangular NW under the transverse electric (TE) polarization. It turns out, that the high-Q modes can be treated as a superposition of TE($m,l$) and TE($m-2,l+2$) or TE($m,l$) and TE($m+2,l-2$) modes accompanied by the avoid-crossing features of real part of the eigenvalues at a given size ratio $R$. Following these general rules, we can immediately find and construct many different high-Q modes. We demonstrate that the Q-factor of a square subwavelength NW can be as high as $2.3 \times 10^4$. The strong confinement of the electric field corresponds to the suppression of the radiation in limited leaky channels or radiation quenching to a minimum in the momentum space. This conclusion can also be generalized to other geometries, such as rectangular NW with transverse magnetic polarization, single cylinder with finite thickness, cuboid, etc. Moreover, we experimentally verify the existence of high-Q modes supported by a single Si NW in the scattering spectrum. Our results will find the applications in boosting the light-matter interaction, such as nonlinear optics effect, strong coupling, and lasers.

The main idea of our approach can be demonstrated in terms of 2×2 Hamiltonian matrix describing a coupling of two modes[5–7]

$$H = \begin{pmatrix} E_1 & V \\ W & E_2 \end{pmatrix} \quad (1)$$

where $E_1$ and $E_2$ are the complex energies of the uncoupled system; V and W are the coupling constants. For an open system, all the parameters are complex, in general, and describe the interaction of leaky modes. For simplicity, we assume $E_1 = E_0 - \gamma_0 i$ and $E_2 = E_0 + \Delta - \gamma_0 i$, where $\gamma_0$ is the radiative rate of two uncoupled modes, and $\Delta$ is the frequency detuning, assumed to be real. The eigenvalue of matrix can be obtained

$$E_\pm(\Delta) = (E_0 + \tfrac{\Delta}{2} - \gamma_0 i) \pm \sqrt{\tfrac{\Delta^2}{4} + VW} \quad (2)$$

In order to realize the infinity large Q-factor (also known as BIC[8–11]), either Im(E+) or Im(E-) should be zero. This condition can be achieved with a complex coupling term, VW. In this case, th square root on the right side is a complex number and can be written as ($s+ti$). The Q-factor for E± can be obtained from Eq.(3) and expressed as

$$Q_\pm(\Delta) = \frac{4 + \Delta \pm 2s}{4(\gamma_0 \mp t)} \quad (3)$$

From Eq.(3), one can find that the Q-factor for E+ approaches to infinity for $t = \pm\gamma_0$, which is the critical condition to form BIC mode.

As an example, Fig.1a-b shows both the real and imaginary parts of $E_\pm$ for $E_0=2$, $\gamma_0=0.01$ and $VW=2\times10^{-4}i$. The avoid-crossing of two modes is present due to the strong coupling condition. At the same time, -Im($E_+$) reduced to zero while -Im($E_-$) reaches its maximum at $\Delta=0$, which means that the critical condition of BIC state is satisfied. Moreover, it is worth noting that VW can be used to modulate the Q-factor. as shown Fig.1c for $E_+$. When the coupling strength deviates from the critical condition, the Q-factor becomes finite and decreases with the reduced value of VW.

However, the maximum still occurs at Δ=0. The high-Q mode can be regarded as a quasi–BIC(QBIC)[12]. Since the maximum Q-factor for a particular case always happens at Δ=0, we can derive the relationship between coupling strength VW=$qi$ and radiative rate $\gamma_0$ for uncoupled modes for both BIC and quasi BIC cases

$$q = 2A^2\gamma_0^2, \quad Q_+(\Delta = 0) = \frac{2+A\gamma_0}{2(1-A)\gamma_0} \quad (4)$$

where the parameter A characterises the degree of deviation from an ideal BIC state. For example, when A=1, Eq.(4) represents the case of BIC with infinity large Q-factor, and when A<1, the high-Q mode corresponds to quasi-BIC. Both cases are presented in Fig.1d, which shows VW and $Q^{-1}$ as functions of the decay rate $\gamma_0$ for A=1 and A=0.8. Therefore, ultra-high Q mode (BIC or QBIC) can be achieved by carefully engineering the coupling strength between two resonances. We have to mention that the conclusion drawn from such a simple case can also be applied to other systems the complex couplings and different decay rates (See Fig.S1-3). In the follows, we will demonstrate that strong coupling condition can be satisfied by changing the size ratio of rectangular NW (or cubic and finite cylinder), and high-Q factor optical nanocavity can be realized at certain size ratio.

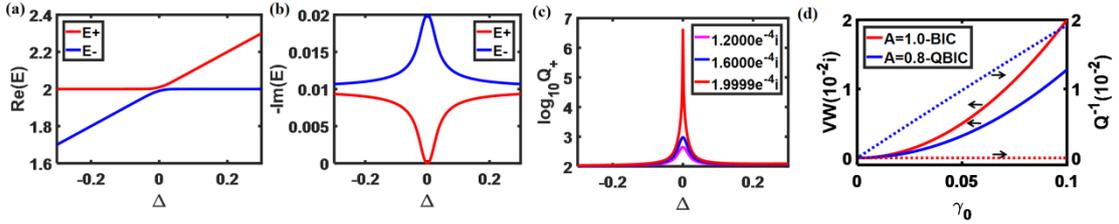

**Fig.1**| BIC and QBIC mode in the two-level system. **a-b,** Real and imaginary part of eigenvalues E± versus Δ for E1=2-0.01i and E2=2+Δ-0.01i and VW=2.0e-4i. **c,** $\log_{10}Q$ of mode E+ versus Δ. **d,** Coupling strength VW and $Q_+^{-1}$ versus $\gamma_0$ when the critical condition is satisfied.

We first consider the eigenmodes (named leaky modes) of a rectangular NW with refractive index $n=4$ under the transverse electric polarization (TE) with electric field along the z-direction. Assuming the width and height of nanowire are *a* and *b*, respectively. The size ratio of NW is defined as $R=b/a$. The leaky modes can be described by the complex eigenvalues $N=n\omega b/c=N_{real}-iN_{imag}$, where ω is the complex eigenfrequency of the leaky mode and *c* is the speed of light. It allows to express the Q-factor in the following from $Q=N_{real}/(2\times N_{imag})$. In previous work, we have demonstrated that the eigenmode supported by the dielectric nanostructure plays the dominant role in describing its optical properties[13–15]. Moreover, linear dependence between $N_{real}$ and the size ratio R has been shown for modes TE(*m,l*), where *m* and *l* correspond to the number of peaks of the electric field within the NW in x and y dimensions. In the following section, we will demonstrate that high-Q and low-Q modes only appear at the critical ratio of NW for a pair of modes TE(*m,l*) and TE(*m-2,l+2*) or TE(*m,l*) and TE(*m+2,l-2*), which suggests that the parity of two modes must be same The pair modes here are divided into four categories(See Table.S1): (1) Type I: *l=m+2*; (2) Type II: *m≤l<m+2*; (3) Type III: *l>m+2*; (4) Type IV: *l<m*. Fig.2a-b shows $N_{real}$ and Q-factor as a function of the size ratio *R* for mode TE(3,5) and TE(5,3) which belongs to Type I. Interestingly, the Q-factor reaches its maximum value 3300 at $R$=1 for TE(3,5) while the avoid-crossing occurs for $N_{real}$ in these two modes. Other high-Q modes fall within the category of Type I, such as TE(1,3) and TE(2,4), can be found at the same critical ratio R=1 (See Fig.S4). Fig.2c-d shows the Q-factor and a/λ (or ka/2π) for mode TE($m,m$+2) while the value of m increases from 1 to 5. The Q-factor can be up to $2.3\times10^4$ for mode TE(5,7) while the resonant wavelength is still larger than the width of square NW. Even higher Q-factor can be obtained for TE($m,m$+2) with $m$>5. For instance, for $m$=6 the Q-factor can get up to $2.98\times10^5$. The resonant wavelength, however, will become smaller than the width of square NW. Thus, there is a balance between the high-Q and the dimensions of the structure. Another interesting point is that a/λ shows a linear dependence on m. Such a linear relationship can be explained from the ray optics perspective (See Section 1 in SM and Fig.S5)[6,16]. It can help to facilitate the process of finding modes even with higher Q-factor for large $m$. Following the same approach, many high-Q modes belonging to Type II amd III can be found (See Fig.S6-10). For the Type IV, note that the complex eigenvalue $N=n\omega b/c$ of TE ($m,l$) for size ratio $R=b/a$ is the same as the eigenvalue $N=n\omega a/c$ for TE($l,m$) for the size ratio $R=a/b$(Fig.S9). Therefore, if Type II or Type III modes pair TE ($l, m$) and TE ($l$+2, $m$-2) display avoid-crossing features at critical ratio R and the Q-factor of TE($l, m$) has maximum value, TE ($m, l$) and TE($m$-2,$l$+2) will show avoid-crossing at 1/R and TE($m,l$) will have the largest Q-factor.

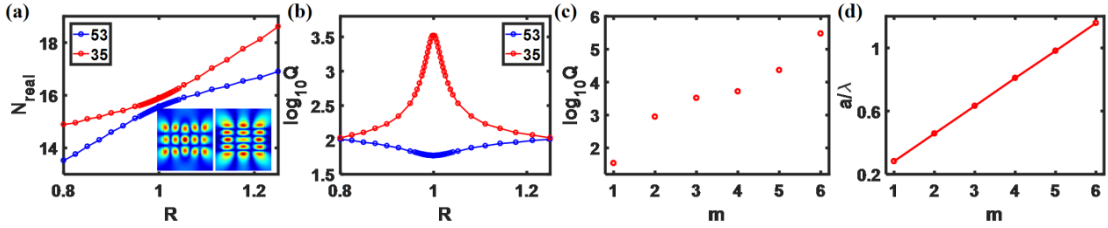

**Fig.2|** Properties of the High-Q modes. **a-b,** Real part and Q-factor of the eigenvalue of modes TE(3,5) and TE(5,3) versus Ratio. **c-d,** Q-factor and a/λ versus m for high-Q mode TE($m,m$+2) at the critical ratio.

We confirm the existence of such high-Q modes at the critical size ratio by calculating the energy density mapping and scattering efficiency mapping versus both the size ratio and the normalized frequency *ka* (See Fig.S11-12). To gain a deep physical insight into radiative properties of the high-Q modes, we employ the multipole expansion method[17,18]. Here, we consider the case of NW at oblique incidence ($\theta$=15°) with TE polarization. Fig.3a shows scattering efficiency contributed by multipoles for square NW. Two resonant peaks can be observed at *ka*=3.89 and *ka*=3.97, which are related to the low-Q mode TE(5,3) and high-Q mode TE(3,5). The scattering efficiency around *ka*=3.97 is dominated by electric quadrupole (*m*=2), exhibiting sharp Fano profile[19,20]. In contrast, there are two dominant multipoles for the low-Q mode around *ka*=3.89. Each multipole can be considered as an independent channel for the radiating decay. These are again confirmed by the multipole analysis on the eigenfield of two modes in Fig.3b. Thus, by coupling to more radiating channels will, in general, reduce the Q-factor. That is why it is expected that high-Q modes should couple to only one radiative channel, described by a single multipole. Importantly, we can also find these radiation channels by carefully comparing its eigenfield distribution to the electric field distribution of the eigenmode for an infinite cylinder (See Fig.S13). Moreover, the radiation intensity for both channels in TE(3,5) is much lower the counterparts in TE(5,3). From the perspective of radiation channel, we may attribute the extreme confinement of TE(3,5) to the fact that these two leaky channels are more confined comparing to leaky channels of TE (5,3). This explanation also works for all of other high-Q modes (See FigsS14-17). Ideally, the eigenmode with the closest field profile to the eigenmode $TE_{ml}$($m>1$ and $l=1$) in the cylinder will always have high Q-factor because there is only one leaky channel with minimum radiation intensity(See Fig.S16-17).

The radiative properties of an arbitrary source can be also analyzed in the momentum space. It is known that only the nonzero current "on-the-shell" in the k-space contributes to the far-field radiation[21]. Thus, it is instructive to analyze the electric field in the momentum space of high-Q mode at the critical ratio to get a deeper physical insight. To do this, we perform the Fourier transform of the eigenfield of the high-Q modes, shown in Fig.3c. Here, it is worth pointing out that the electric field E($k_0$) contributes to the outward radiation only when $k_x^2+ k_y^2=k_0^2$. Therefore, we extract E($k_0$) on the white circle boundary $k_x^2+ k_y^2=k_0^2$, and plot them in Fig.3d. Indeed, the radiation field E($k_0$) of high-Q mode has much lower amplitude, and the radiation channel is narrower. For the resonant mode with extremely high Q-factor, E($k_0$) approaches zero at the circle boundary ($k_x^2+ k_y^2=k_0^2$).

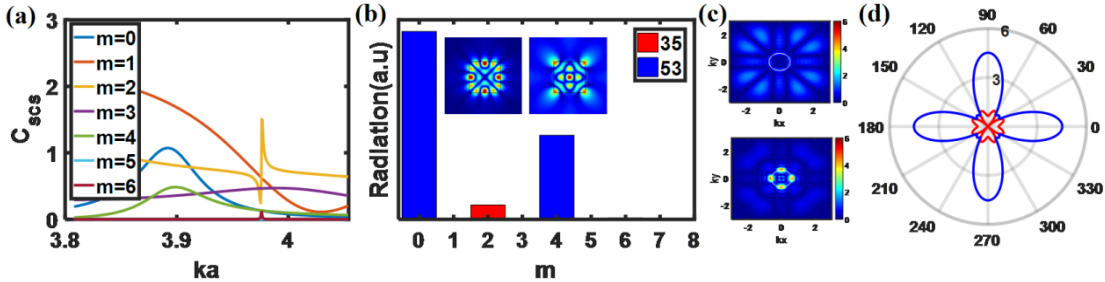

**Fig.3| Multipole analysis of the high-Q modes. a,** Multipolar contribution on the scattering cross section of the square NW with $\theta_{inc}$=15°. **b-c,** Multipole analysis and Fourier transformation on the eigenfields of two modes. **d,** E($k_0$) obtained from Fourier transformation of eigenfields for two modes.

The above phenomenon can also be generalized to the rectangular NW with TM polarization (See Fig.S18-23). As described in two level system, coupling between these two modes is the prerequisite to achieve high-Q, we calculate the coupling strength for both TE and TM cases[22,23](See Section 3 in SM,Table.S2 and Fig.S24-26). It is also noted that such high Q modes exist for structures with different refractive index (See Fig.S27).

So far, we only discuss how to find the quasi BIC induced high-Q mode for a single rectangular NW. The above approach can also be applied to a three-dimensional nonspherical structure including the cuboid and cylinder with finite thickness. Here, we use a single cuboid as an example to demonstrate how to find a high-Q mode. For the sake of convenience, $a=b$ and $R=c/a$ are assumed. Also, the mode number along x is chosen as 1 for simplicity. Only magnetic eigenmodes M(1,2,3) and M(1,4,1) in the yOz plane are investigated here as an example. Remarkably, the Q-factor reaches a maximum of 325 around avoid corssing(R=0.795). The results of multiple analysis on the modes M(1,4,1) and M(1,2,3) indicate that main radiation channels are $l=1$ and $l=3$, respectively. Careful examination on the eigenfield distribution tells us that these two channels can be linked to eigenmodes $M_{12}$ and $M_{31}$ for sphere nanoparticle. $E(k_0)$ obtained from Fourier transform on eigenfield has much lower amplitude for M(1,2,3) than that of mode M(1,4,1). This again confirms that the radiation is suppressed to a minimum. More high-Q modes for single cuboid and disk can be found using similar way (See Fig.S28-31).

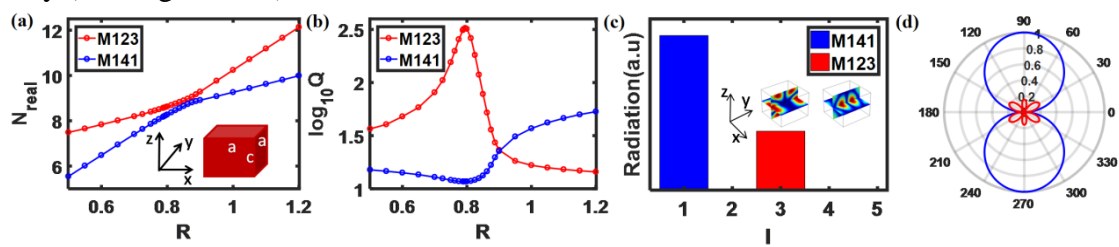

**Fig.4|** High-Q modes of a single cuboid. a-b, $N_{real}$ **(a)** and Q factor **(b)** versus size ratio for leaky modes M(1,2,3) and TM(1,4,1). **c,** Multipole analysis on the eigenfields for two modes. d. $E(k_0)$ distribution for two modes.

Finally, we demonstrate the high-Q factor by exploring the scattering characteristics of single rectangular silicon NW on a quartz substrate. Mode TE(3,4) is used to realize the high-Q mode at 1390nm. The height of the silicon NW is 825nm. A series of NW with different width are defined by electron beam lithography. Figs.5b-c show the measured scattering spectrum of a single nanowire with different size ratio. Good agreement can be found between numerical calculation based on FEM method and experimental results (See Fig.S32). The resonant frequency and Q-factor are extracted by standing fano fitting procedure(See Section 2 in SM and Fig.S32)[19]. Indeed, the quality factor can be up to 211 for a single NW around 1390nm at R=0.868. Note also, that the resonant frequency matches very well to the theoretical prediction. A slight shift of the critical size ratio may be attributed to the imperfect vertical sidewall of NW. Note that the quality factor can be further improved to about 746 by suspending the NW, which can be realized by wet etching $SiO_2$ underneath using HF acid. Furthermore, high-Q mode for TM case is also investigated for the same structure. Besides, we also demonstrate that the Q-factor for mode TE(3,5) can reach 380 for silicon NW on quartz substrate with thickness being 1130nm (See Fig.S33-34) while Q-factor can be up to 294 for TM(3,5) (See Fig.S35).

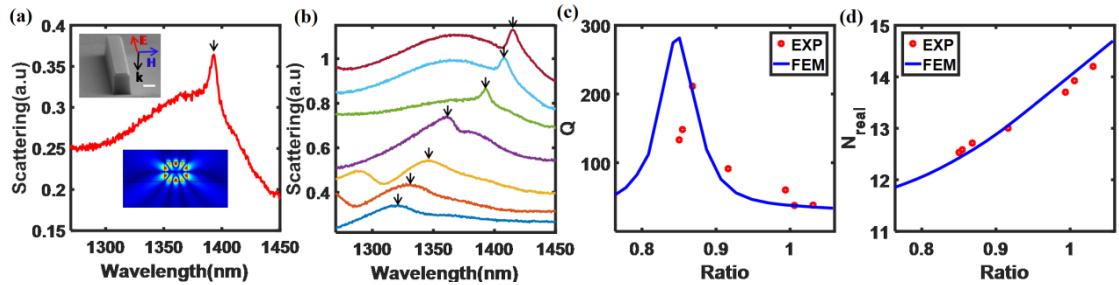

Fig.5 Experimental verification of high-Q mode in single silicon NW. a, Measured scattering spectrum for single NW with a=950nm and b=825nm at normal incidence for TE polarization. The insets are SEM image for fabricated Si NW on quartz and eigenfield distribution. b, Measured scattering spectrum for single NW with different width for b=825nm. c-d, Extracted Q factor (c) and $N_{real}$ (d) versus size ratio.

In summary, we report the upper limit of the high-Q modes for a single dielectric nonspherical nanocavity by exploring quasi-BIC. We also experimentally confirm this type of high-Q ( ≈380) mode in single Si subwavelength NW. Our findings provide a general guiding principle to design extreme high-Q mode with relatively small material volume and will find the applications in lasing with lowthreshold, biosensor and enhancing light-matter interactions.

## Acknowledgements
This research was supported by the Australian Research Council (ARC) and the UNSW Scientia Fellowship program.

## References
1. Akahane, Y., Asano, T., Song, B.-S. & Noda, S. High-Q photonic nanocavity in a two-dimensional photonic crystal. *Nature* **425**, 944–947 (2003).
2. Vahala, K. J. Optical microcavities. *Nature* **424**, 839 (2003).
3. Kuznetsov, A. I., Miroshnichenko, A. E., Brongersma, M. L., Kivshar, Y. S. & Luk'yanchuk, B. Optically resonant dielectric nanostructures. *Science (80-. ).* **354**, aag2472 (2016).
4. Kivshar, Y. & Miroshnichenko, A. Meta- Optics with Mie Resonances. 24–31 (2017).
5. Cao, H. & Wiersig, J. Dielectric microcavities: Model systems for wave chaos and non-Hermitian physics. *Rev. Mod. Phys.* **87**, 61–111 (2015).
6. Wiersig, J. Formation of long-lived, scarlike modes near avoided resonance crossings in optical microcavities. *Phys. Rev. Lett.* **97**, 1–4 (2006).
7. Song, Q. H. & Cao, H. Improving optical confinement in nanostructures via external mode coupling. *Phys. Rev. Lett.* **105**, 2–5 (2010).
8. Hsu, C. W. *et al.* Observation of trapped light within the radiation continuum. *Nature* **499**, 188 (2013).
9. Hsu, C. W., Zhen, B., Stone, A. D., Joannopoulos, J. D. & Soljačić, M. Bound states in the continuum. *Nat. Rev. Mater.* **1**, 16048 (2016).
10. Kodigala, A. *et al.* Lasing action from photonic bound states in continuum. *Nature* **541**, 196

(2017).

11. Ha, S. T. *et al.* Directional lasing in resonant semiconductor nanoantenna arrays. *Nat. Nanotechnol.* **13**, 1042–1047 (2018).
12. Rybin, M. V *et al.* High-$Q$ Supercavity Modes in Subwavelength Dielectric Resonators. *Phys. Rev. Lett.* **119**, 243901 (2017).
13. Huang, L., Yu, Y. & Cao, L. General Modal Properties of Optical Resonances in Subwavelength Nonspherical Dielectric Structures. *Nano Lett.* **13**, 3559–3565 (2013).
14. Yu, Y. & Cao, L. Coupled leaky mode theory for light absorption in 2D, 1D, and 0D semiconductor nanostructures. *Opt. Express* **20**, 13847–13856 (2012).
15. Cao, L. *et al.* Engineering light absorption in semiconductor nanowire devices. *Nat. Mater.* **8**, 643 (2009).
16. Wiersig, J. Hexagonal dielectric resonators and microcrystal lasers. *Phys. Rev. A - At. Mol. Opt. Phys.* **67**, 12 (2003).
17. Li, S.-Q. & Crozier, K. B. Origin of the anapole condition as revealed by a simple expansion beyond the toroidal multipole. *Phys. Rev. B* **97**, 245423 (2018).
18. Grahn, P., Shevchenko, A. & Kaivola, M. Electromagnetic multipole theory for optical nanomaterials. *New J. Phys.* **14**, (2012).
19. Miroshnichenko, A. E., Flach, S. & Kivshar, Y. S. Fano resonances in nanoscale structures. *Rev. Mod. Phys.* **82**, 2257–2298 (2010).
20. Tribelsky, M. I. & Miroshnichenko, A. E. Giant in-particle field concentration and Fano resonances at light scattering by high-refractive-index particles. *Phys. Rev. A* **93**, 53837 (2016).
21. Kim, K. & Wolf, E. Non-radiating monochromatic sources and their fields. *Opt. Commun.* **59**, 1–6 (1986).
22. Bogdanov, A. A. *et al.* Bound states in the continuum and Fano resonances in the strong mode coupling regime. *Adv. Photonics* **1**, 1–12 (2019).
23. Zhang, L., Gogna, R., Burg, W., Tutuc, E. & Deng, H. Photonic-crystal exciton-polaritons in monolayer semiconductors. *Nat. Commun.* **9**, 713 (2018).